\documentclass[twocolumn,showpacs]{revtex4}
\usepackage[dvips]{graphicx}

\def\D#1#2{\frac{\partial #1}{\partial #2}}
\def\d#1#2{\frac{d\, #1}{d\, #2}}

\begin{document}
\title{The Quantum Big Bang in Global Time Theory}
\author{D.E. Burlankov}
\email{bur@phys.unn.runnet.ru}
\affiliation{Physics Department, University of Nizhny Novgorod, Russia.}

\begin{abstract}
The {\it Global Time Theory} (GTT) is the further development of the General
Relativity (GR). GTT  significantly differs from GR in the general
physical concepts, but retains 90\% of the mathematical structure and main results.
The dynamics equations are derived from Lagrangian, and the Hamiltonian of gravitation is nonzero.
The quantum theory of gravitation can be built on the basis of the Schroedinger
equation, as for other fields. The quantum model of the Big Bang is demonstrated.
\end{abstract}

\pacs{04, 04.60.-m, 98.80.Hw}
\maketitle

\section{Global Time Theory}\

The Global Time Theory (GTT) (see \cite{Bjern}) is the further development of General Relativity (GR).
GTT significantly
differs  from GR in base physical postulates, but retains 90\% of the mathematical structure and main results.
Meanwhile, description of the cosmos dynamics in GTT leads to significant modifications.

In GTT time is absolute. It flows equally -- always and everywhere -- and is itself the
measure of equality. The development of {\it the Entire Universe} occurs in this global time.

The space has three dimension, is Riemanian, and its metric tensor ($\gamma_{ij}$)
can depend on space coordinates and time. Points of space are linked with it absolutely. The
frame of reference in which coordinates of space points do not change is called {\it
the global inertial system}.

The inertial system permits arbitrary tree-dimensional transformations of coordinates $\tilde{x}^i(x^j)$
(that do not depend on time).

The coordinate transformations that are {\it  time-dependent}  lead to {\it global non-inertial
system} of observer. Meanwhile the time remain global. In the non-inertial system
the {\it vector field of absolute velocities} $V^i$ arises, although it vanish in the
inertial system.

Thus there are nine components of fields, described the space dynamics: six components
of three-metric $\gamma_{ij}$ and three components of  absolute velocities $V^i$.

The space dynamics describes in global time but usually in {\it non-inertial  system}
which coordinates $x^i$ are linked with inertial coordinates $\bar{x}^i$ by
transformations, depending on the global time $t$. As sequence of this transformations
the time derivative in inertial system is expressed through the derivatives in the
non-inertial system by the {it covariant derivative over the time}.
The structure of  covariant derivatives over the time of tensors acquires additions
in the form of Lie-variation, that are generated by transformation of coordinates  $\delta x^i=-V^i\,dt$ --
for return  to the inertial system.

For a tensor of an arbitrary range
 \begin{equation}
D_t\:Q^i_{jk}=\D{}{t}Q^i_{jk}-V^i_{;s}Q^s_{jk}+V^s_{;j}Q^i_{sk}+
V^s_{;k}Q^i_{js}+V^sQ^i_{jk;s}. \label{covtens}
\end{equation}

For scalar there is
\begin{equation}
D_t\:F=\D{F}{t}+\D{F}{x^i}\:\D{x^i}{t}=\D{F}{t}+V^i\D{F}{x^i},
\label{covtscal}
\end{equation}
which determines the covariant derivative over the time of scalar field (action, eikonal) in
an arbitrary frame with global time.

Especially important for the theory is the covariant derivative over the time of the
metric tensor:
\begin{equation}
D_t\:\gamma _{ij}=\D{\gamma _{ij}}{t}+V_{i;j}+V_{j;i}.
\label{difmtr}
\end{equation}

\subsection{Action and dynamical equations}\

In GR, space play a rather passive role.
In GTT the three-dimensional space  is the dynamic field, relative to which there
exists an absolute motion, or, on the contrary, there exists a field of space
velocities in a given system of coordinates. The equations of motion
are derived from the variational principle. Lagrangian is presented as the difference
of the kinetic and potential energy.
Introducing the {\it tensor of space deformation velocity}
\begin{equation}
\mu _{ij}=\frac{1}{2\,c}D_t\,\gamma _{ij}=\frac{1}{2\,c}(\dot{\gamma}_{ij}+
V_{i;j}+V_{j;i}), \label{muij}
\end{equation}
we can represent this action as preconcerted with the Hilbert action in GR:
\begin{equation}
 S=\frac{c^4}{16\,\pi\,k}\int(\mu^i_j\,\mu^j_i-(\mu^j_j)^2+R)
\sqrt{\gamma}\,d_3\,x\,dt+ S_m,
\label{Lagr}
\end{equation}
where $S_m$ is the action of enclosed matter, which adds to dynamic equations the
energy-momenta tensor components. The absolute velocities $V^i$ are represented only
in kinetic energy.

By introducing momenta
$$\pi^i_j=\frac{\sqrt{\gamma}}{2}\,(\mu^i_j-\delta^i_j\,\mu^s_s),$$
and varying the action by six components of spacial metrics, we obtain the six
equations of dynamics:
\begin{equation}\label{qij}
\dot{\pi}^i_j=b^i_j+\sqrt{\gamma}\,G^i_j+\sqrt{\gamma}\,(T^i_j-V^i\,T^0_j),
\end{equation}
where $b^i_j$ we call {\it the self tensor current}
\begin{equation}\label{self}
 b^i_j=-\delta^i_j\,\frac{1}{\sqrt{\gamma}}(2\,\pi^k_l\pi^l_k-\pi^k_k\pi^l_l)-
\partial_s(V^s\,\pi^i_j)+V^i,_s\,\pi^s_j-V^s,_j\,\pi^i_s,
\end{equation}
$G^i_j$ is the Einstein's tensor of three-dimensional space, and $T^\alpha_\beta$ --
components of tensor energy-momenta of enclosed matter, which
determine {\it the exterior tensor current}.

In three-dimensional space, the Riemann-Kristoffel tensor is algebraically expressed
through the Einstein's tensor ($\epsilon_{ijk}$ -- absolute antisymmetric tensor):
$$R^{ij}_{kl}=-\epsilon^{ijs}\epsilon_{klm}G^m_s,$$
and therefore the absence of currents leads to a flat space.

The variation (only the kinetic part of action) by three components of the field of
absolute velocities gives three equations of links:
\begin{equation}\label{varV}
  \nabla_i\,\pi^i_j=\sqrt{\gamma}\,\frac{8\pi k}{c^4}T^0_j.
\end{equation}

The Hamiltonian :
$$H=\int \pi^{ij}\dot{\gamma}_{ij}\,d_3x-L=$$
\begin{equation}\label{HamGr}
\int\left(\frac{c^4}{16\pi k}\left(\frac{\pi^i_j\pi^j_i-(\pi^i_i)^2/2}{\sqrt{\gamma}}-
R\,\sqrt{\gamma}\right)-2\,\pi^i_j\,V^j_{;i})\right)\,d_3x.
\end{equation}
Its unique feature is the non fixed sign, as a result of which such phenomena as
Friedman cosmological expansion are possible.

\subsection{The proper time of the moving body}\

Similar to GR, the GTT encompasses special relativity. On level with global time,
in which the
growth of Entire Universe occur, for the moving observer there are his local system
and {\it local time}. All local phenomena in the moving system develop in this local time.
This can be expressed  through the square of absolute velocity
\begin{equation}\label{loctime}
d\tau=dt\,\sqrt{1-\gamma_{ij}(\dot{x}^i-V^i)(\dot{x}^j-V^j)}.
\end{equation}

This expression can be represented in four-dimensions by joining the time and
space into unified four-dimensional manifold with metrics
\begin{equation}\label{fourmtr}
  g_{00}=1-\gamma_{ij}\,V^i\,V^j;\quad g_{0i}=\gamma_{ij}\,V^j;\quad
  g_{ij}=-\gamma_{ij}.
\end{equation}
The reverse metric tensor of this four-dimensional manifold is
$$g^{00}=1;\quad g^{0i}=V^i;\quad g^{ij}=V^i\,V^j-\gamma^{ij}.$$

The first equation carries great significance
\begin{equation}\label{Glob}
g^{00}=1.
\end{equation}
This is {\it the main structural relationship} in GTT, analogous to the Minkowski
metric, which is the main structural relationship in special relativity.

\subsection{General Relativity}\

If there is the four-dimensional metric $g_{\alpha\beta}$ in arbitrary four coordinates
$x^\alpha,\,\alpha=0..3$, the variable $\tau$ must be determined for reduction to
global time, in order for the main structure relation (\ref{Glob}) to hold true. We
must transfer the metrics component $g^{00}$ by rule of tensor transformation:
\begin{equation}\label{HJGTT}
  \bar{g}^{00}=g^{\alpha\beta}\D{\tau}{x^\alpha}\D{\tau}{x^\beta}=1.
\end{equation}
This differential equation turns out to be Hamilton-Jacoby differential equation for free-falling bodies
(laboratories), the common time for which is $\tau$, which is the global time.

Thus {\it the equivalence principle} is realized, but in contrast to GR, the time of
the inertial system exists not only for a local laboratory, but for a great many
laboratories in all of space.

GTT does not mathematically coincide with GR only in one equation:
since the main structure relation (\ref{Glob})  $g^{00}=1$ not permit to vary this
component, the tenth Einstein's equation, determined by this component variation
in GTT is absent. This tents equation of GR in GTT represents as
\begin{equation}\label{zeroH}
  H=0,
\end{equation}
where $H$ is the common density of the Hamiltonian of space and enclosed matter.

Thus the GR solusions is the subset of GTT solutions with particular value of energy
equal to zero.

\section{Big Bang in GTT}

So as the Hamiltonian in GTT is not equal zero, the effective
quantum theory of gravitation can be built on the basis of the Schroedinger
equation, as for other fields.

\subsection{Classical solutions}\

Now we study the compact cosmological model of Friedmanian type with space as a
three-dimensional sphere with variable radius $r$, depended on the time $t$.
This Universe  is filled by ultra-relativistic matter with the state equation
$\varepsilon=3\,p$.

The Lagrangian of isoenthrophic gas id expressed by integral over space of pressure,
determined as function of the chemical potential \cite {Schutz} $\mu$ as function (in homogeneous case)
of the time derivative of {\it inner action}: $\mu=\dot{\sigma}$:
\begin{equation}\label{LagGaz}
  L_m=2\pi^2\int\, p({\mu})\,r^3.
\end{equation}

The action variation by $\sigma$ yields the matter conservation law
\begin{equation}\label{ConsMat}
  \d{}{t}(\rho\,r^3)=0;\quad\rho=\d{p}{\mu}.
\end{equation}

For ultrarelativistic matter the pressure is proportional $\mu^4$,
what together with Lagrangian of space yields the full Lagrangian:
\begin{equation}\label{LagCosm}
L=-\frac{3\,\pi\,c^2}{4\,k}\,r\,\dot{r}^2+\frac{3\,\pi\,c^4\,r}{4\,k}+\frac{\dot{\sigma}^4}{4\,a^6}\,r^3,
\end{equation}
where constant $a$ has dimension of length.

Further we will works in almost Plankean system of units where light velocity $c=1$,
$2\,k/(3\,\pi)=1$ and $\hbar=1$. All physical values are non-dimensional and the
dimensional energy $E$ express through non-dimensional $e$ by expression
$$E=e\,c^2\sqrt{\frac{3\,\pi\,c\,\hbar}{2\,k}}.$$

The non-dimensional Lagrangian is

$$L=\frac{r\,(1-\dot{r}^2)}{2}+\frac{\dot{\sigma}^4}{4\,a^6}\,r^3.$$
The momentum conjugated to $\sigma$ is constant
$$p_r=-r\,\dot{r};\quad p_\sigma=\frac{\dot{\sigma}^3\,r^3}{a^6}.$$

By expression the velocities through momenta
$$\dot{r}=-\frac{p_r}{r};\quad\dot{\sigma}=\frac{1}{r}\left(\frac{p_\sigma}{\rho_0}\right)^{1/3},$$
we can express the Hamiltonian
\begin{equation}\label{Hcosm}
  H=-\dot{r}\,p_r+\dot{\sigma}\,p_\sigma-L=-\frac{p_r^2+r^2}{2\,r}+\frac{q^2}{2\,r},
\end{equation}
where $q=\sqrt{3/2}\,a\,p_\sigma^{(2/3)}$ and determines the conserved quantity of the ultrarelativistic matter.

The classical equation of motion can be derived from the energy conservation law:
\begin{equation}\label{eqFrRel}
  \dot{r}=\frac{1}{r}\,\sqrt{q^2-2E\,r-r^2}.
\end{equation}

This equation describes the radius oscillations between the maximal and minimal
values, which are determines by roots of the subroot expression.

\begin{equation}\label{rmax}
r_{max}=\sqrt{e^2+q^2}-e.
\end{equation}

The second root is negstive.

If $q^2\neq 0$, the energy can be as negative as positive.
At $q^2=0$ -- clear gravitational dynamics without matter -- the energy can j,tain
only negative values,

\subsection{Quantum model}\

The wave function is function of the radius $r$.
By $u'$ we denote the derivative of the wave function $u(r)$ over $r$.
By symmetryzation  the product  $p^2/r$ we gets the cosmological wave equations:
$$-\left(\frac{u'}{r}\right)'+\left(-r+\frac{q^2}{r}\right)\,u=2\,E\,u.$$

\begin{equation}\label{qeq}
 u''-\frac{u'}{r}+(-r^2+q^2)\,u=2\,r\,E\,u:
\end{equation}

This equation has a regular critical point $r=0$ and irregualar $r=\infty$, at that
vicinity the wave function behave oneself as the ones of oscillator:
$$u(r)\sim A\,e^{-r^2/2}+B\,e^{r^2/2}.$$

At the some values of $E$ the coefficient $B$ vanishes -- there are normalizable
solutions of the quantum equation. The function is equal zero at $r=0$ and infinity,
and consequently it can have $n$ extremums.

At the vicinity of zero radius all solutions behave oneself as $r^2$. This means that
probability density  at $r=0$ in any  state is zero.

The equation (\ref{qeq}) have two parameters: $q$ and $e$.

The self energy values for small $n$ at $q^2=0,\,1,\,10$
are represented in next table

\vspace{10pt}
\begin{center}

\begin{tabular}{|c|c|c|c|} \hline
n&$q^2=0$&$q^2=1$&$q^2=10$\\
\hline
  1 & -1.3133& -1.0202&2.6765 \\
  2 &-1.9243 & -1.7122&0.564 \\
  3 &-2.3863 & -2.2107&-0.441 \\
  4 &-2.773  &-2.6193 &-1.1208\\
  \hline
  \dots&\dots&\dots&\dots\\
  \hline
  8&-3.9599&-3.8487&-2.8153\\
  \hline
\end{tabular}
\end{center}

The eigenvalues of energy can be equal to zero (as in GR), but only at the $q^2=4\,n$.
At this value of $q$ there is solely  wave function and only with $n$ extremums.

This model demonstrate the difference between quantum and classical solutions. The
solutions of the classical equation  (\ref{eqFrRel}) describes the radius oscillations
between the positive maximal and negative minimal radiuses.
The point $r=0$ in classical solution is usual point. In quantum solutions the point
$r=0$ is critical point and the value of the wave function at this point is zero, so
the quantum oscillations of radius take place between zero and maximal classical radius.

\section{Conclusion}\

The main philosophical difference of GTT from GR is in its rejection of the {\it
principle of general covariance}. This difference is not in its mathematical treatment
(upon coordinate transformation all equations must be transformed correctly),
but in its physical vulgarization, i.e the confirmation of the principal
{\it physical} equality of all nonsingular coordinate systems, and as a result --
the confirmation of principal nonpossibility of the
introduction of global time. GTT revives the global time, which simplifies and
enriches the theory.

The main {\it physical} difference of GTT from GT is in nonzero energy density. In
contrast to GR, in GTT the quantum theory of gravitation is built on the basis of the Schroedinger
equation, as for other fields.

Author thanks Ksenia P. Brazhnik for help in translation.

\end{document}